# Discrete Persistent-Chain Model for Protein Binding on DNA

Pui-Man Lam[1] and Yi Zhen[2]
Physics Department, Southern University
Baton Rouge, Louisiana 70813

We describe and solve a discrete persistent chain model of protein binding on DNA, involving an extra $s_i$ at a site i of the DNA. This variable takes the value 1 or 0 depending on whether the site is occupied by a protein or not. In addition, if the site is occupied by a protein, there is an extra energy cost $e$. For small force, we obtain analytic expressions for the force-extension curve and the fraction of bound protein on the DNA. For higher forces, the model can be solved numerically to obtain force extension curves and the average fraction of bound proteins as a function of applied force. Our model can be used to analyze experimental force extension curves of protein binding on DNA, and hence deduce the number of bound proteins in the case of non-specific binding.

PACS numbers:87.14gk,82.35.Rs,87.15.kj,61.30.Hn

## I.    Introduction

It is well known that binding of proteins on DNA may enhance (or reduce) the effective persistence length of DNA. In the packaging of centimeter size DNA into microscopic size chromosomes, the persistence length of DNA is very much reduced. We cannot think of an experimental situation where the persistence length of DNA is enhanced by protein but in principle this is also possible. Therefore we will also investigate this situation theoretically. Protein binding on DNA also plays a fundamental role in many cellular and viral functions, including gene expression, physical chromosome organizations, chromosome replication, and genetic recombination [1-3]. Therefore understanding the effect of protein binding on DNA is an important topic in cell biology. Micromanipulation techniques using magnetic tweezers are important tools for such studies [4-9], by measuring the extension versus force characteristics of single molecules of DNA in the presence of fixed amounts of protein in solution.

In such experiments, it is of interest to determine the number of bound proteins on the DNA. When the protein binds to specific DNA targets of defined sequence, this can be done relatively easily, e.g. by observation of jumps in apparent DNA extension at some fixed force [4, 10, 11], or by detection of fluorescently labeled proteins [12]. However, the situation is more problematic when the protein binding on the DNA is nonspecific, i.e., it can bind to essentially any site along a long DNA strand. To attack this problem, Zhang and Marko [8] applied analogs of the Maxwell relations of classical thermodynamics to determine changes in numbers of bound proteins, from measurements of extension as a function of bulk protein concentration. This method has been applied by Liebesny et al [9] to analyze their extension versus force measurements of $\lambda$ DNA in the presence of $\lambda$ repressor protein (CI).

In this paper we present an alternate, more microscopic approach that compliments Zhang and Marko's thermodynamic method. For this purpose we generalize the discrete persistent-chain model of Storm and Nelson [13, 14] to include interaction with the protein. We show that within this model, we can derive analytic expressions for the force-extension curve and the number of proteins as a function of force. We find that in the case when the protein reduces (enhances) the effective persistence length of the DNA, the number of bound proteins decreases (increases) with the external applied force.

In section II we present a detailed description of the model and its solution. Section III is a summary.

## II.    Discrete Persistent Chain Model with Protein Binding

The energy of the model of DNA with N monomers, interacting in solution with protein, is given by [13]

$$\frac{E[\{\hat{t}_i\}]}{kT} = \sum_{i=1}^{N}\left[\frac{A_0}{b}(1-\hat{t}_i\cdot\hat{t}_{i+1})d_{s_i,0} + \frac{A_1}{b}(1-\hat{t}_i\cdot\hat{t}_{i+1})d_{s_i,1} - \frac{fb}{kT}\hat{t}_i\cdot\hat{z} + es_i\right]. \quad (1)$$

Here $A_0$, $A_1$ are the persistent lengths with and without proteins respectively. b is the Kuhn length and $\hat{t}_i$ is the unit tangent vector along the DNA at the i-th site. $s_i$ is a variable at the i-th site, taking the value of either unity or zero, depending on whether that site contains a bound protein or not. A site without any bound protein has a persistent length $A_0$, while that with a bound protein has a persistent length $A_1$. The first two terms are the energy cost, proportional to the corresponding persistent length at that site, for two consecutive unit tangent vectors not pointing in the same direction,. f is the external pulling force along the z-direction, applied at one end of the DNA, k is the Boltzmann's constant and T is the temperature. $e$ is the energy cost for binding a protein on the DNA. For $e = \infty$, the model reduces to the original discrete persistent-chain model without protein, while for $e = -\infty$, every site of the DNA is bound with a protein.

The partition function is given by

$$Z = \sum_{\{s_i\}}\left[\prod_{i=1}^{N}\int_{S^2}d^2\hat{t}_i\right]\exp\left(-\frac{E[\{\hat{t}_i\}]}{kT}\right), \quad (2)$$

where $\int_{S^2}d^2\hat{t}_i$ means an integration over the unit sphere of the i-th unit tangent vector.

Substituting Eqn. (1) into Eqn. (2) and defining $l_0 = A_0/b$, $l_1 = A_1/b$, and $\tilde{f} = \frac{fb}{kT}$, we have

$$Z = \left[\prod_{i=1}^{N}\int_{S^2} d^2\hat{t}_i\right]\sum_{s_1=0}^{1}\exp\left(-(1-\hat{t}_1\cdot\hat{t}_2)(l_0 d_{s_1,0}+l_1 d_{s_1,1})-\mathbf{e} s_1+\tilde{f}\hat{t}_1\cdot\hat{z}\right)$$

$$\times \sum_{s_2=0}^{1}\exp\left(-(1-\hat{t}_2\cdot\hat{t}_3)(l_0 d_{s_2,0}+l_1 d_{s_2,1})-\mathbf{e} s_2+\tilde{f}\hat{t}_2\cdot\hat{z}\right)\times\bullet\bullet\bullet$$

$$=\left[\prod_{i=1}^{N}\int_{S^2} d^2\hat{t}_i\right]\left[\exp\left(-(1-\hat{t}_1\cdot\hat{t}_2)l_0+\tilde{f}\hat{t}_1\cdot\hat{z}\right)+\exp\left(-(1-\hat{t}_1\cdot\hat{t}_2)l_1+\tilde{f}\hat{t}_1\cdot\hat{z}-\mathbf{e}\right)\right]$$

$$\times\left[\exp\left(-(1-\hat{t}_2\cdot\hat{t}_3)l_0+\tilde{f}\hat{t}_2\cdot\hat{z}\right)+\exp\left(-(1-\hat{t}_2\cdot\hat{t}_3)l_1+\tilde{f}\hat{t}_2\cdot\hat{z}-\mathbf{e}\right)\right]\bullet\bullet\bullet$$

$$=\left[\prod_{i=1}^{N}\int_{S^2} d^2\hat{t}_i\right]\exp\left(\frac{1}{2}\tilde{f}\hat{t}_1\cdot\hat{z}\right)\left[\exp\left(\frac{1}{2}\tilde{f}(\hat{t}_1+\hat{t}_2)\cdot\hat{z}\right)\left(\exp[-(1-\hat{t}_1\cdot\hat{t}_2)l_0]+\exp[-(1-\hat{t}_1\cdot\hat{t}_2)l_1-\mathbf{e}]\right)\right]$$

$$\times\left[\exp\left(\frac{1}{2}\tilde{f}(\hat{t}_2+\hat{t}_3)\cdot\hat{z}\right)\left(\exp[-(1-\hat{t}_2\cdot\hat{t}_3)l_0]+\exp[-(1-\hat{t}_2\cdot\hat{t}_3)l_1-\mathbf{e}]\right)\right]\times\bullet\bullet\bullet$$

$$\times\left[\exp\left(\frac{1}{2}\tilde{f}(\hat{t}_{N-1}+\hat{t}_N)\cdot\hat{z}\right)\left(\exp[-(1-\hat{t}_{N-1}\cdot\hat{t}_N)l_0]+\exp[-(1-\hat{t}_{N-1}\cdot\hat{t}_N)l_1-\mathbf{e}]\right)\right]\exp\left(\frac{1}{2}\tilde{f}\hat{t}_N\cdot\hat{z}\right)$$

(3)

From this we can define a transfer matrix $\Im$ with matrix elements

$$\Im(\hat{t}_i,\hat{t}_j) = \exp\left(\frac{1}{2}\tilde{f}(\hat{t}_i+\hat{t}_j)\cdot\hat{z}\right)\left(\exp[-(1-\hat{t}_i\cdot\hat{t}_j)l_0]+\exp[-(1-\hat{t}_i\cdot\hat{t}_j)l_1-\mathbf{e}]\right), \quad (4)$$

with matrix multiplication defined as

$$\Im(\hat{t}_i,\hat{t}_j)v(\hat{t}_i) \equiv \int_{S^2} d^2\hat{t}_j\,\Im(\hat{t}_i,\hat{t}_j)v(\hat{t}_j) \quad , \quad (5)$$

where $v(\hat{t}_i)$ is any function of the unit tangent vector $\hat{t}_i$. Then neglecting the boundary terms and in the limit $N\to\infty$, the partition function can be written as

$$Z = \mathbf{1}_{\max}{}^N , \tag{6}$$

where $\mathbf{1}_{\max}$ is the maximum eigenvalue of the transfer matrix $\mathfrak{I}$. From this it follows that the average extension $z$ along the direction of the force and the average number $M$ of bound protein is given by

$$\left\langle \frac{z}{L} \right\rangle = \frac{\partial}{\partial \tilde{f}} \ln \mathbf{1}_{\max} \tag{7}$$

$$\langle m \rangle = \left\langle \frac{M}{N} \right\rangle = \frac{\partial}{\partial \boldsymbol{e}} \ln \mathbf{1}_{\max} , \tag{8}$$

where L is the total length of the DNA.

In order to calculate the maximum eigenvalue $\mathbf{1}_{\max}$, we will follow the procedure of Storm and Nelson [13]. We define a one parameter family of trial functions $v_{\boldsymbol{w}}(\hat{t})$ as

$$v_{\boldsymbol{w}}(\hat{t}) = \exp(\boldsymbol{w}\hat{t} \cdot \hat{z}) , \tag{9}$$

with magnitude squared given by

$$\|v_{\boldsymbol{w}}\|^2 = \int_{S^2} d^2\hat{t} \exp(2\boldsymbol{w}\hat{t} \cdot \hat{z}) = \frac{2p}{\boldsymbol{w}} \sinh(2\boldsymbol{w}) \tag{10}$$

Then the maximum eigenvalue $\lambda_{\max}$ can be variationally approximated by

$$\mathbf{1}_{\max}^* = \max y(\boldsymbol{w}) \equiv \max \frac{v_{\boldsymbol{w}} \cdot \mathfrak{I} \cdot v_{\boldsymbol{w}}}{\|v_{\boldsymbol{w}}\|^2} , \tag{11}$$

where we have to maximize with respect to $\boldsymbol{w}$.

$$\|v_{\boldsymbol{w}}\|^2 y(\boldsymbol{w}) = \exp(-l_0) \int_{S^2} d^2\hat{t}_i \int_{S^2} d^2\hat{t}_{i+1} \exp\left(\frac{1}{2}(\tilde{f} + 2\boldsymbol{w})(\hat{t}_i + \hat{t}_j) \cdot \hat{z} + \hat{t}_i \cdot \hat{t}_j l_0\right)$$

$$+ \exp(-l_1 - \boldsymbol{e}) \int_{S^2} d^2\hat{t}_i \int_{S^2} d^2\hat{t}_{i+1} \exp\left(\frac{1}{2}(\tilde{f} + 2\boldsymbol{w})(\hat{t}_i + \hat{t}_j) \cdot \hat{z} + \hat{t}_i \cdot \hat{t}_j l_1\right) \tag{12}$$

Define the function $H(\mathbf{w},\tilde{f},l)$ as

$$H(\mathbf{w},\tilde{f},l) \equiv \frac{\mathbf{w}}{2\pi \sinh(2\mathbf{w})} \exp(-l) \int_{S^2} d^2\hat{t}_i \int_{S^2} d^2\hat{t}_{i+1} \exp\left(\frac{1}{2}(\tilde{f}+2\mathbf{w})(\hat{t}_i+\hat{t}_j)\cdot\hat{z} + \hat{t}_i\cdot\hat{t}_j l\right). \tag{13}$$

Then we can write $y(\mathbf{w})$ as

$$y(\mathbf{w}) = H(\mathbf{w},\tilde{f},l_0) + e^{-e} H(\mathbf{w},\tilde{f},l_1) \tag{14}$$

The function $H(\mathbf{w},\tilde{f},l)$ can be calculated explicitly. It has the form [14]

$$H(\mathbf{w},\tilde{f},l) = \frac{8\pi \mathbf{w}\, \mathrm{cosech}(2\mathbf{w})}{l(\tilde{f}+2\mathbf{w})} \exp\left[-\frac{3}{2}l - \frac{(\tilde{f}+2\mathbf{w})^2}{8l}\right] \int_{|l-(\tilde{f}/2+\mathbf{w})|}^{l+\tilde{f}/2+\mathbf{w}} dG \exp[G^2/(2l)]\sinh(G) \tag{15}$$

This integral can be evaluated in terms of error function. The result is

$$H(\mathbf{w},\tilde{f},l) = \frac{-i2\sqrt{2}\pi^{3/2}\mathbf{w}}{\sqrt{l}(2\mathbf{w}+\tilde{f})} \exp\left[-2l - \frac{(2\mathbf{w}+\tilde{f})^2}{8l}\right] \mathrm{cosech}(2\mathbf{w})$$
$$\times \left\{\Phi\left(i\frac{4l+\tilde{f}+2\mathbf{w}}{2\sqrt{2l}}\right) - \Phi\left(i\frac{4l-(\tilde{f}+2\mathbf{w})}{2\sqrt{2l}}\right) - 2\Phi\left(i\frac{\tilde{f}+2\mathbf{w}}{2\sqrt{2l}}\right)\right\}, \tag{16}$$

where the error function is defined as

$$\Phi(x) = \frac{2}{\sqrt{\pi}} \int_0^x e^{-t^2} dt \tag{17}$$

Expanding Eqn. (16) to second order in $\mathbf{w}$ and $\tilde{f}$, we find

$$H(\mathbf{w},\tilde{f},l) = c(l)\left(1 - \frac{2}{3}\mathbf{w}^2 + \frac{1}{12}(\tilde{f}+2\mathbf{w})^2[1+\Lambda(l)]\right), \tag{18}$$

where

$$\Lambda(x) = \coth(x) - \frac{1}{x} \tag{19}$$

is the well known Langevin function and the function $c(x)$ is defined as

$$c(x) = \frac{4p}{x} e^{-x} \sinh(x) \qquad (20)$$

Substituting Eqn. (16) into Eqn. (14) for $y(\mathbf{w})$ and expanding $y(\mathbf{w})$ to second order in $\omega$ and f one finds

$$y(\mathbf{w}) = c_0 + c_1 e^{-e} + \frac{1}{3}[c_0(1+\Lambda_0) + c_1 e^{-e}(1+\Lambda_1)]w\tilde{f}$$

$$-\frac{w^2}{3}[c_0(1-\Lambda_0) + c_0 e^{-e}(1-\Lambda_1)] + \frac{\tilde{f}^2}{12}[c_0(1+\Lambda_0) + c_1 e^{-e}(1+\Lambda_1)], \qquad (21)$$

where $\Lambda_0 = \Lambda(l_0)$, $\Lambda_1 = \Lambda(l_1)$ and $c_0 = c(l_0)$, $c_1 = c(l_1)$.

From Eqn. (21) one can analytically calculate $\omega^*$ that maximizes $y(\mathbf{w})$ to be

$$\mathbf{w}^* = \frac{\tilde{f}}{2} \frac{c_0(1+\Lambda_0) + c_1 e^{-e}(1+\Lambda_1)}{c_1(1-\Lambda_0) + c_1 e^{-e}(1-\Lambda_1)}, \qquad (22)$$

Substituting Eqn. (22) into Eqn. (21) one finds the maximum eigenvalue to be

$$\mathbf{1}_{max} = y(\mathbf{w}^*) = (c_0 + c_1 e^{-e})\left(1 + \frac{\tilde{f}^2}{6} \frac{c_0(1+\Lambda_0) + c_1 e^{-e}(1+\Lambda_1)}{c_0(1-\Lambda_0) + c_1 e^{-e}(1-\Lambda_1)}\right) \qquad (23)$$

From this one can obtain the average extension as

$$\left\langle \frac{z}{L} \right\rangle = \frac{\partial}{\partial \tilde{f}} \ln \mathbf{1}_{max}$$

$$= \frac{\tilde{f}}{3} \frac{c_0(1+\Lambda_0) + c_1 e^{-e}(1+\Lambda_1)}{c_0(1-\Lambda_0) + c_1 e^{-e}(1-\Lambda_1) + \frac{f^2}{6}[c_0(1+\Lambda_0) + c_1 e^{-e}(1+\Lambda_1)]}. \qquad (24)$$

The average number of bound proteins is given by

$$\langle m \rangle = \left\langle \frac{M}{N} \right\rangle = \frac{\partial}{\partial e} \ln \mathbf{1}_{max}$$

$$= \frac{c_1 e^{-e}}{c_0 + c_1 e^{-e}} + \frac{\tilde{f}^2}{6} \frac{c_0 e^{-e}[(c_0-c_1)(1-\Lambda_0\Lambda_1) - (c_0+c_1)(\Lambda_0-\Lambda_1)] + c_1 e^{-2e}(c_0-c_1)(1-\Lambda_1^2)}{[c_0(1-\Lambda_0) + c_1 e^{-e}(1-\Lambda_1)]^2}$$

. (25)

Since the Langevin function $\Lambda(x)$ is a monotonic increasing function of $x$ and is always less than unity, and $c(x)$ is a monotonic decreasing function of $x$ for $x > 1$, we can see that for $l_0 > l_1$, i.e. when the bound protein decreases the persistent length of the DNA, the numerator in the second term of Eqn. (25) is negative and so $\langle m \rangle$ is a decreasing function of $f$. For $l_0 < l_1$, its sign changes to positive and hence $\langle m \rangle$ becomes an increasing function of $f$. Also it can be seen from Eqn. (25) that the average fraction of bound protein at zero external applied force is given exactly by the expression

$$< m_0 >= \frac{c_1 e^{-e}}{c_0 + c_1 e^{-e}} \qquad (26)$$

Expressions (24) and (25) are correct only for small values of $f$. For general values of $f$, one first obtains analytic expressions for $\frac{\partial \ln y}{\partial \tilde{f}}$, $\frac{\partial \ln y}{\partial e}$ and $\frac{\partial y}{\partial w}$ by using Eqn. (14). Then for fixed values of $\tilde{f}$ and $e$ the optimal value $w^*$ can be obtained numerically by solving $\frac{\partial y(w)}{\partial w} = 0$. Substituting this value of $w^*$ into the expressions for $\frac{\partial \ln y}{\partial \tilde{f}}$ and $\frac{\partial \ln y}{\partial e}$ we obtain the average extension and number of bound proteins for these values of $\tilde{f}$ and $e$. In Fig. 1a we show the average extension versus force for the case $l_0 = 50$, $l_1 = 25$, for different values of $e$, the energy cost in binding a protein with the DNA. We see that the average extension decreases with decreasing $e$, compared with the case of no protein, which corresponds to the case of $e = \infty$. In Fig. 1b we show the average number of bound proteins $< m >$ versus the applied force, for the same values of $l_0$ and $l_1$. In this case we find that $< m >$ is a decreasing function of the force. In Fig. 2 we show the average fraction of bound protein $< m >$ as a function of $e$ for $\tilde{f} = 0.1$. The average fraction of bound protein $< m >$ is unity for large negative $e$ and becomes zero for large positive $e$. In Fig. 3a we show the average extension versus force for the case $l_0 = 50$, $l_1 = 100$. In this case we find that extension increases with decreasing $e$. In Fig. 3b we show the average number of bound proteins $< m >$ versus the applied force, for these values of $l_0$ and $l_1$. In this case we find that $< m >$ is an increasing function of the force.

### III. Summary and Conclusion

We have given a detailed description of a discrete persistent chain model of protein binding on a DNA. The proteins are described by a variable $s_i$ at a site i of the DNA. This variable takes the value 1 or 0 depending on whether the site is occupied by a protein or not. If the site is occupied by a protein, there is an extra energy cost $e$. This variable can be summed exactly in the partition function and the remaining variable $\hat{t}_i$ that are unit tangent vectors at the site i can be summed using transfer matrix technique, originally employed by Storm and Nelson [13]. This results in an analytic expression for the partition function, involving the error function. For small force, the partition function can be expanded and we obtain analytic expressions for the force-extension curve and the fraction of bound protein on the DNA. From these, we can see that in the case when the protein reduces (enhances) the effective persistence length of the DNA, the number of bound proteins decreases (increases) with the external applied force. When no external force is applied, we obtain an exact expression for the fraction of bound proteins, involving the effective persistent lengths $l_0$ and $l_1$ with or without proteins respectively, and the energy cost $e$ for the binding of a protein with DNA. For higher forces, the model can be solved numerically to obtain force extension curves and the average fraction of bound proteins as a function of applied force. Our model can be used to analyze experimental force extension curves of protein binding on DNA, and hence deduce the number of bound proteins in the case of non-specific binding.

Our model is fundamentally different from that of Zhang-Marko. This can be most easily seen by considering the case of vanishing external force $f$. When the external force $f=0$, the Zhang-Marko model gives the same values for the average fraction of bound proteins $<m>$ at fixed values of $bm$, which corresponds to $-e$ in our model, independent of the values of the persistent lengths $A_0$ and $A_1$ for the DNA with and without protein respectively. The reason for this is because the formula (10) in the Zhang-Marko paper for $<m>$ depends on $A_0$ and $A_1$ only through the functions $g(f)$ and $h(f)$ which are both zero for $f=0$. In our model, the value of $<m>$ still depends on the persistent lengths $A_0$ and $A_1$ even when $f=0$. This is because our Hamiltonian, equation (1) still depends on $A_0$ and $A_1$ even when $f=0$ and consequently so does the average $<s_i>$. The average $<m>$ at zero external force is given correctly by our equation (26). In particular, for $e=0$, the average $<m>$ at $f=0$ still depends on $A_0$ and $A_1$, contrary to the value 1/2 predicted by Zhang-Marko model for $f=0$ and $bm=0$. Also in our model, we have assumed that the proteins can bind only to the sites of the base-pairs.

Note added: After our paper was submitted, we learned that the effect of protein on DNA was also studied using the discrete worm like chain model, analyzed by the spherical harmonics expansion [15,16].

**Acknowledgement.** Research supported by the Louisiana Board of Regents Support Fund Contract Number LEQSF(2007-10)-RD-A-29 and Louisiana Epscor EPS-1003897 and NSF(2010-15)-RII-SUBR. PML acknowledge the hospitality of the Institute of Theoretical Physics, Chinese Academy of Sciences, where part of this work was carried out.

1 **puiman_lam@subr.edu**
2 **yi_zhen@subr.edu**

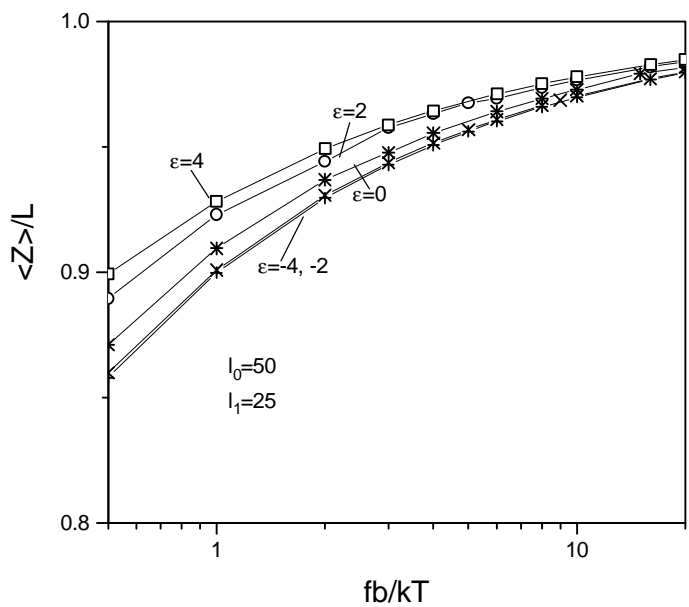

Fig. 1a: The average extension $\left\langle \dfrac{z}{L} \right\rangle$ as a function of $\tilde{f}$ for the case $l_0 = 50$, $l_1 = 25$, for different values of *e*.

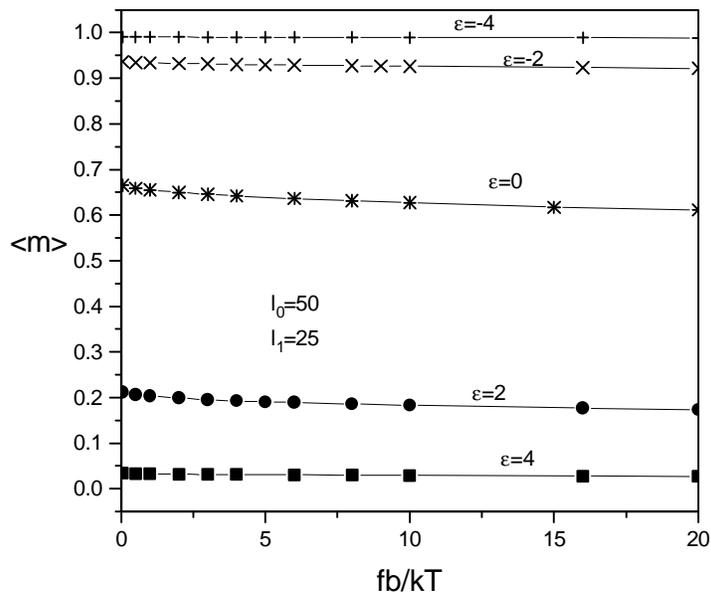

Fig. 1b: The average number of bound proteins $<m>$ as a function of $\tilde{f}$ for the case $l_0 = 50$, $l_1 = 25$, for different values of $e$.

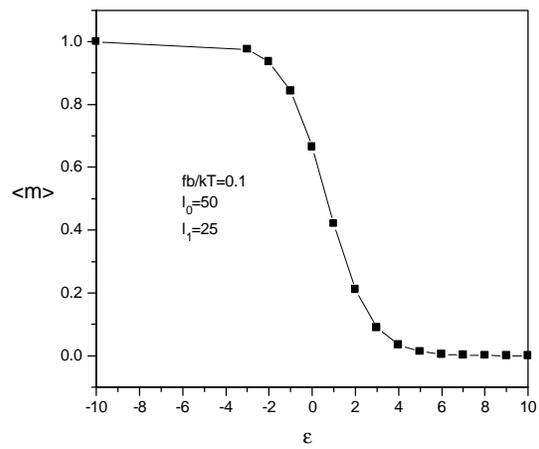

Fig. 2: Average fraction of bound protein as a function of $e$, for the case $l_0 = 50$, $l_1 = 25$

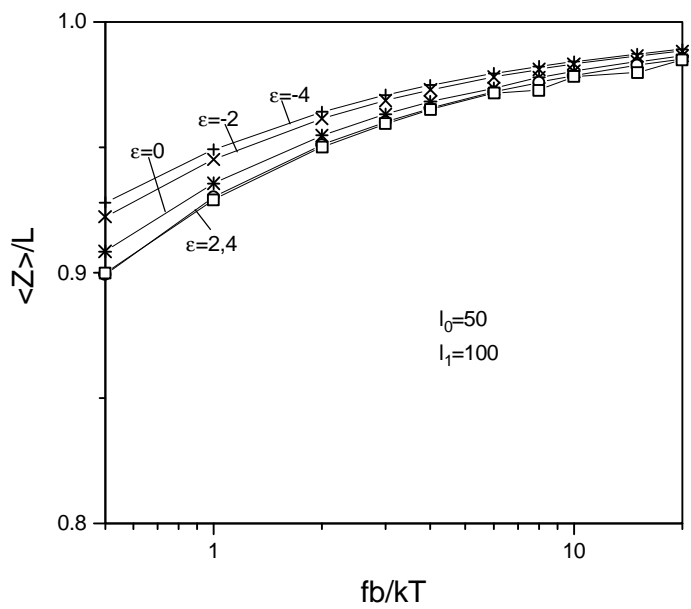

Fig. 3a: The average extension $\left\langle \dfrac{z}{L} \right\rangle$ as a function of $\tilde{f}$ for the case $l_0 = 50$, $l_1 = 100$, for different values of $e$.

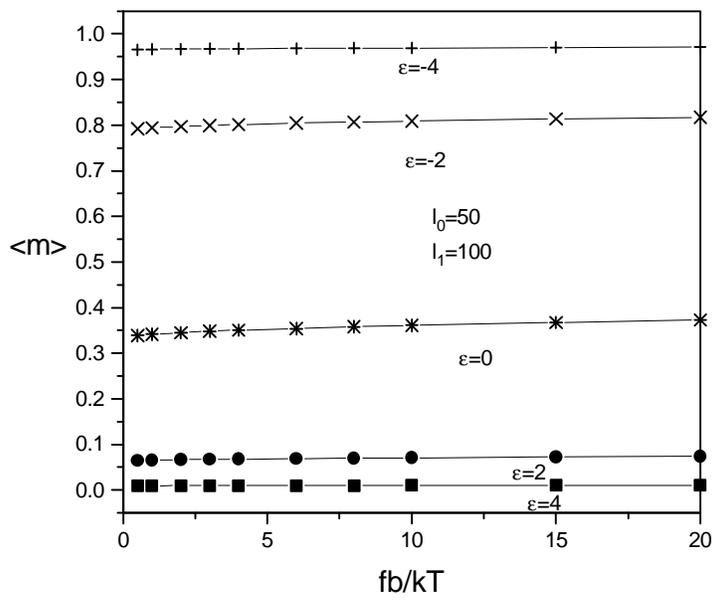

Fig. 3b: The average number of bound proteins $<m>$ as a function of $\tilde{f}$ for the case $l_0 = 50$, $l_1 = 100$, for different values of $\varepsilon$.

**References**

[1] J.R. Broach, Cell **119**, 583 (2004)
[2] R.G. Roeder, Nat. Med. **9**, 1239 (2003)
[3] B. Alberts, D. Bray, A. Johnson, N. Lewis, K. Roberts, P. Walter and M. Raff, *Essential Cell Biology: An Introduction to the Molecular Biology of the Cell* (Garland Publishing Inc., New York, U.S.A., 1997)
[4] L. Finzi and J. Gelles, Science **267**, 378 (1995)
[5] J. F. Marko and E.D. Siggia, Biophys. J. **73**, 2173 (1997)
[6] S. Cocco, J.F. Marko, R. Monasson, A. Sarkar and J. Yan, Eur. Phys. J. E **10**, 249 (2003)
[7] J. Yan and J.F. Marko, Phys. Rev. E **68**, 011905 (2003)
[8] H. Zhang and J.F. Marko, Phys. Rev. E **77**, 031916 (2008)
[9] P. Liebesny, S. Goyal, D. Dunlap, F. Family and L. Finzi, J. Phys.: Condensed Matter **22,** 414104(2010)
[10] G. Lia, D. Bensimon, V. Croquette, J.F. Alleman, D. Dunlap, D.E. Lewis, S. Adhya and L. Finzi, Proc. Natl. Acad. Sci. U.S.A. **100**, 11373 (2003)
[11] S.F. Tolic-Norrelykke, M.B. Rasmussen, F.S. Pavone, K. Berg-Sorensen and L.B. Oddershede, Biophys. J. **90**, 3694 (2006)
[12] J. Mameren, M. Modesti, R. Kanaar, C. Wyman, G. J. Wuite and E.J. Peterman, Biophys. J. **91**, L78 (2006)
[13] C. Storm and P.C. Nelson, Phys. Rev. E **67**, 051906 (2003)
[14] P.M. Lam, Phys. Rev. E **70**, 013901 (2004)
[15] J. Yan and J.F. Marko, Phys. Rev. **E68**, 011905 (2003)
[16] H. Zhang and J.F. Marko, Phys. Rev. **E82**, 051906 (2010)